\documentclass{elsart}

\usepackage{adnd}
\usepackage{longtable}

\usepackage{mathptmx}

\usepackage{amsmath}

\usepackage{amssymb,amstext}
\usepackage[pdftex,
    letterpaper=true,
    hyperindex=true,
    breaklinks=true,
    colorlinks=false,
    citecolor=blue,
    pdftitle={},
    pdfauthor={}]
{hyperref}

\usepackage{graphicx}

\begin{document}

\begin{frontmatter}

\journal{Atomic Data and Nuclear Data Tables}

\copyrightholder{Elsevier Science}

\runtitle{Arsenic}
\runauthor{Shore}


\title{Discovery of the Arsenic Isotopes}


\author{A.~Shore},
\author{A.~Fritsch},
\author{M.~Heim},
\author{A.~Schuh},
\and
\author{M.~Thoennessen\corauthref{cor}}\corauth[cor]{Corresponding author.}\ead{thoennessen@nscl.msu.edu}

\address{National Superconducting Cyclotron Laboratory and \\ Department of Physics and Astronomy, Michigan State University, \\East Lansing, MI 48824, USA}

\date{1/9/2009} 

\begin{abstract}
Twenty-nine arsenic isotopes have so far been observed; the discovery of these isotopes is discussed.  For each isotope a brief summary of the first refereed publication, including the production and identification method, is presented.
\end{abstract}

\end{frontmatter}





\newpage
\tableofcontents
\listofDtables

\vskip5pc

\section{Introduction}\label{s:intro}

The second paper in the series of the discovery of isotopes \cite{Gin09}, the discovery of the arsenic isotopes is discussed. In the growing amount of information gathered for more and more isotopes, the basic knowledge of where and how the isotopes were first produced tends to get lost. The purpose of this series is to document and summarize the discovery of the isotopes. Guidelines for assigning credit for discovery are (1) clear identification, either through decay-curves and relationships to other known isotopes, particle or $\gamma$-ray spectra, or unique mass and Z-identification, and (2) publication of the discovery in a refereed journal. The authors and year of the first publication, the laboratory where the isotopes were produced as well as the production and identification methods are discussed. When appropriate, references to conference proceedings, internal reports, and theses are included. When a discovery included a half-life measurement the measured value is compared to the currently adapted value taken from the NUBASE evaluation \cite{Aud03} which is based on the ENSDF database \cite{ENS08}. In cases where the reported half-life differed significantly from the adapted half-life (up to approximately a factor of two), we searched the subsequent literature for indications that the measurement was erroneous. If that was not the case we credited the authors with the discovery in spite of the inaccurate half-life.

\section{Discovery of $^{64-92}$As}
Presented in this article are the 29 discovered isotopes of arsenic, from A = $64-92$. There is one stable, 11 proton-rich and 17 neutron-rich isotopes. The HFB-14 model predicts $^{110}$As and $^{117}$As to be the heaviest even and odd particle-bound isotope of arsenic, respectively \cite{Gor07}. For neutron-deficient isotopes, the proton dripline has been crossed. There is potentially one more unbound isotope ($^{63}$As) that is estimated to live long enough to be observed \cite{Tho04}. Thus 23 isotopes or about 45\% of all arsenic isotopes remain undiscovered.

\begin{figure}
	\centering
	\includegraphics[width=12cm]{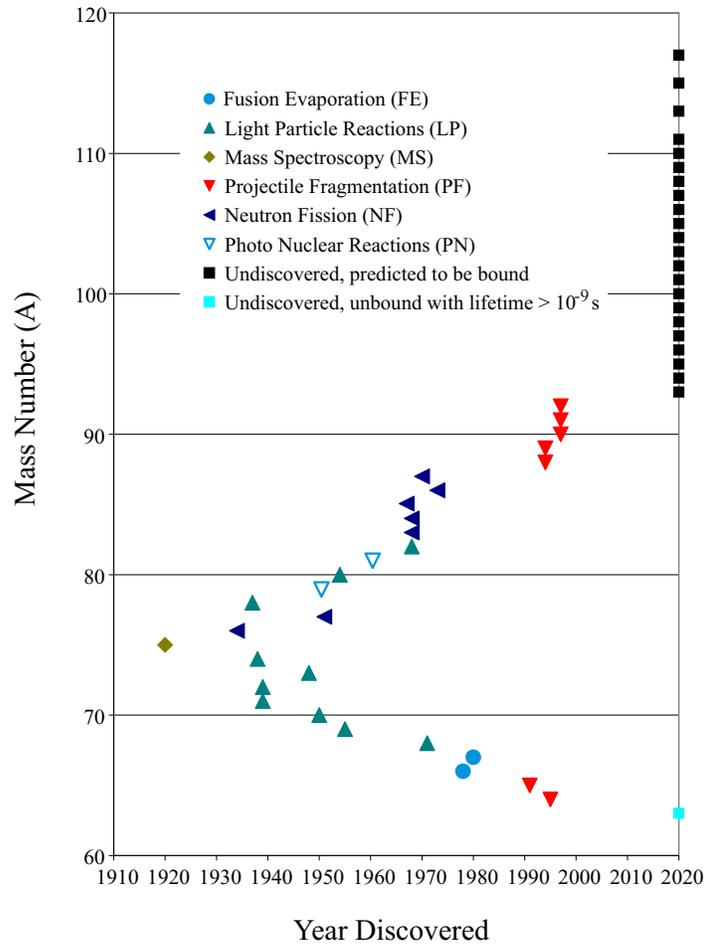}
	\caption{Arsenic isotopes as a function of time they were discovered. The different production methods are indicated. The solid black squares on the right hand side of the plot are isotopes predicted to be bound by the HFB-14 model.  On the proton-rich side the light blue square corresponds to an unbound isotope predicted to have a lifetime larger than $\sim 10^{-9}$~s.}
	\label{f:year}
\end{figure}

Figure \ref{f:year} depicts the year of discovery for each arsenic isotope as identified by the production method. The stable isotope $^{75}$As was discovered by mass spectroscopy (MS). The production methods to produce the radioactive isotopes were light-particle reactions (LP), photo-nuclear reactions (PN), neutron-induced fission (NF), fusion-evaporation (FE) and projectile fragmentation or projectile fission (PF). Heavy ions are all nuclei with an atomic mass larger than A~=~4 \cite{Gru77}. Light particles also include neutrons produced by accelerators. In the following the discovery of each arsenic isotope is discussed in detail.

\subsection*{$^{64}$As}\vspace{-0.85cm}
In the 1995 article \textit{New Isotopes from $^{78}$Kr Fragmentation and the Ending Point of the Astrophysical Rapid-Proton-Capture Process} Blank \textit{et al.} reported the discovery of $^{64}$As at the SISSI/LISE facility of the Grand Acc\'{e}l\'{e}rateur National d'Ions Lourds in Caen, France, via the projectile fragmentation of a 73~MeV/nucleon $^{78}$Kr beam on a nickel target \cite{Bla95}. The new isotope was identified by its time of flight through the separator and the $\Delta$\textit{E}-\textit{E} in a silicon detector telescope. A lower limit for the half-life was established, ``The observation of $^{64}$As in our experiment and the comparison of the counting rate to neighboring nuclei excludes half-lives much shorter than 1~$\mu$s.''

\subsection*{$^{65}$As}\vspace{-0.85cm}
Mohar \textit{et al}. first observed $^{65}$As via projectile fragmentation in 1991 at the National Superconducting Cyclotron Laboratory at Michigan State University: \textit{Identification of New Nuclei near the Proton-Dripline for 31$\leq$Z$\leq$38} \cite{Moh91}. A 65~A$\cdot$MeV $^{78}$Kr beam produced by the K1200 cyclotron reacted with a $^{58}$Ni target. $^{65}$As was identified by measuring the separator rigidity, $\Delta$\textit{E}, \textit{E}$_{total}$, and the ion velocity. ``The newly commissioned A1200 beam-analysis device was used to observe the astrophsically interesting isotope $^{65}$As.''

\subsection*{$^{66}$As}\vspace{-0.85cm}
 Alburger discovered $^{66}$As at Brookhaven National Laboratory in 1978 via the fusion-evaporation reaction $^{58}$Ni($^{10}$B,2\textit{n}) as reported in \textit{Half-Lives of $^{62}$Ga, $^{66}$As, and $^{70}$Br} \cite{Alb78}. A 30~MeV beam from the MP Tandem Van de Graaff accelerator irradiated the target and $\beta$-rays were detected with a scintillator. ``Seven runs were made on the $^{66}$As half-life at $\beta$ biases from 3.5-4.5~MeV ... the half-life obtained for $^{66}$As was 95.78$\pm$0.39~ms.'' This value is included in the average value of 95.77(23)~ms currently accepted. The authors do not claim credit for the discovery by quoting a previously unpublished report which had measured the correct half-life (93(5)~ms) \cite{Jac76}.

\subsection*{$^{67}$As}\vspace{-0.85cm}
In the 1980 article \textit{$\beta$$^+$ Decay of $^{67}$As} Murphy \textit{et al.} reported the discovery of $^{67}$As at Argonne National Laboratory \cite{Mur80}. The isotope was produced via the fusion-evaporation reaction $^{58}$Ni($^{14}$N,$\alpha$\textit{n}) where the $^{14}$N ions were accelerated by the FN tandem accelerator to energies of 39 and 41~MeV. ``The half life, decay scheme, and mass excess of $^{67}$As have been determined from $\beta$-delayed $\gamma$-ray singles and $\gamma$-$\gamma$ coincidence, and $\beta$-$\gamma$ coincidence measurements.'' The half-life was found to be 42.5(12)~s which is the currently accepted and only available value for $^{67}$As. It should be mentioned that Murphy \textit{et al.} already reported the results of the experiment in a conference abstract in 1976 \cite{Mur76}.

\subsection*{$^{68}$As}\vspace{-0.85cm}

$^{68}$As was first observed by T. Paradellis \textit{et al.} in 1971 as reported in \textit{The Decay of $^{68}$As} \cite{Par71}. A 92\% enriched $^{70}$GeO$_2$ target was bombarded with 40 MeV protons accelerated by the McGill Synchrocyclotron. Characteristic $\gamma$-rays were detected following activation with the $^{70}$Ge(p,3n)$^{68}$As reaction. ``The weighted average of several measurements of the half-life of the 1016.5 keV $\gamma$-ray was found to be 159$\pm$4 sec.'' This half-life is close to the currently accepted value of 151.6(8)~s. The authors do not claim the discovery of $^{68}$As, instead referring to the 1955 paper by Butement and Prout \cite{But55}. However, the cautious statement ``Between 40 and 60~MeV the initial half-life was $\sim$10~minutes, suggesting the formation of a mixture of $^{69}$As with a shorter lived activity, of half-life $\sim$7~minutes, possibly $^{68}$As produced by a (p,3n) reaction on $^{70}$Ge,'' in combination with a significantly deviating half-life value does not warrant credit for the discovery of $^{68}$As by Butement and Prout.

\subsection*{$^{69}$As}\vspace{-0.85cm}
In \textit{Radioactive $^{69}$As and $^{70}$As} Butement and Prout noted their 1955 observation of $^{69}$As at the Atomic Energy Research Establishment in Harwell, England, when samples of germanium dioxide were irradiated with protons \cite{But55}. Mass assignment was made by milking off and identifying the radioactive germanium daughter and the $\gamma$ radiation was measured with a NaI(Tl) scintillator. ``Between 20 and 35~MeV a new 15~minute activity appeared in appreciable yield ($^{69}$As) , due to a (p,2n) reaction on $^{70}$Ge.'' This half-life agrees with the accepted value of 15.2(2) m.

\subsection*{$^{70}$As}\vspace{-0.85cm}
In the 1950 paper \textit{Spallation Products of Arsenic with 190~MeV Deuterons} Hopkins identified the isotope  $^{70}$As \cite{Hop50}. A pure $^{75}$As target was bombarded with 190~MeV deuterons from the Berkeley 184-inch cyclotron and chemically separated and subjected to spectrographic analysis. ``Table 1 contains two changes in isotope assignment differing from those previously reported. The 44-min. selenium and 52-min. arsenic daughter are placed at mass 70 since careful separations revealed no active germanium daughter.'' In a previous paper the activity was incorrectly assigned to $^{71}$As \cite{Hop48}. The observed half-life agrees with the accepted value of 52.6(3)~m.

\subsection*{$^{71}$As}\vspace{-0.85cm}
Sagane discovered $^{71}$As in 1939 at the University of California at Berkeley as reported in \textit{Radioactive Isotopes of Cu, Zn, Ga and Ge} \cite{Sag39}. The Radiation Laboratory cyclotron provided the deuterons that bombarded a germanium target and $^{71}$As was produced in the reaction $^{70}$Ge(d,n). Activities were measured with a Lauritsen-type quartz fiber electroscope. ``The 50-hr. period found in deuteron bombardments is expected to be caused by an arsenic isotope, probably $^{71}$As, because it emits positrons.'' The half life was determined to be 50(3)~hours which is somewhat lower than the accepted value of 65.28(15)~h.

\subsection*{$^{72}$As}\vspace{-0.85cm}
In 1947 Mitchell \textit{et al}. reported in \textit{Radiations from $^{72}$As} the discovery of $^{72}$As at Indiana University when gallium was bombarded with alpha particles from the cyclotron at an energy of 23~MeV followed by chemical separation \cite{Mit47}. ``A preliminary investigation of the radiations from $^{72}$As has been made with the help of the coincidence counting apparatus available in this laboratory ... It is shown to be a positron emitter of approximately 26~hours half-life.'' This half-life is consistent with the accepted value of 26.0(1) h.

\subsection*{$^{73}$As}\vspace{-0.85cm}
McCown \textit{et al}. were the first to correctly identify $^{73}$As in \textit{Radioactive Arsenic Isotopes} at the Ohio State University in 1948 \cite{McC48}. GeO$_{2}$ enriched in isotope 70 and regular GeO$_{2}$ were bombarded simultaneously with alpha-particles. The decay activity was measured with a Wulf Electrometer attached to a freon-filled ionization chamber. ``Since stable $^{70}$Ge was approximately four times as abundant in the enriched $^{70}$Ge sample as in the ordinary Ge, it follows that this 76-day activity is produced from the stable $^{70}$Ge isotope. Hence, the assignment of this activity, produced by the ($\alpha$, \textit{p}) reaction on $^{70}$Ge, is made to $^{73}$As.'' $^{73}$As decays by K-electron capture with a half life of 76(3) d which is consistent with the accepted value of 80.30(6) d. It should be mentioned that in 1939 Sagane had incorrectly attributed an 88(2) m half-life to $^{73}$As \cite{Sag39}.

\subsection*{$^{74}$As}\vspace{-0.85cm}
Sagane \textit{et al.} of the Radiation Laboratory of the University of California at Berkeley were first to observe $^{74}$As as reported in the 1938 article \textit{Radioactive As Isotopes} \cite{Sag38}. Samples of germanium were activated with a 5.5~MeV deuteron beam at the Berkeley cyclotron with the reaction $^{73}$Ge(\textit{d},\textit{n}). In an additional experiment samples of arsenic were bombarded with fast neutrons from the Li + d reaction at the Tokyo cyclotron. ``...we are certain that the process should be written as follows: Ge$^{73}$ + D$^2$ $\rightarrow$ As$^{74}$ + n$^1$ [and] As $^{75}$ + n$^1$ $\rightarrow$ As $^{74}$ + 2n$^1$.'' The half-life was found to be 17~days which is consistent with the accepted value of 17.77(2)~d.

\subsection*{$^{75}$As}\vspace{-0.85cm}
In 1920 Aston identified $^{75}$As at the Cavendish Laboratory in Cambridge in \textit{Mass Spectra and Isotopes} by analyzing the spectrum of the gaseous hydride AsH$_{3}$ using a mass spectrograph \cite{Ast20}. Aston states that the element appears to have no (other) isotopes and in the summary table he assigns it a mass of 75.

\subsection*{$^{76}$As}\vspace{-0.85cm}
In 1934 Amaldi \textit{et al}. discovered $^{76}$As when irradiating targets with neutrons from beryllium powder mixed with emanation (radon) at the Istituo Fisico della R. Universit\`{a} in Rome, Italy, which he announced in \textit{Radioactivity Produced by Neutron Bombardment-V} \cite{Ama34a}. ``A chemical separation of the active substance in presence of gallium and germanium enables us to exclude the possibility that it is gallium and makes it very unlikely that it is germanium. The most probable hypothesis is that the activity is due to $^{76}$As.'' $\beta$-ray activity was measured with a Geiger-M\"{u}ller counter. In a separate paper a half-life of ``about 2 days'' \cite{Ama34b} was determined. The accepted half-life is 1.0778(20)~d.

\subsection*{$^{77}$As}\vspace{-0.85cm}
 The discovery of $^{77}$As was published in 1951 by Steinberg and Engelkemeir from Argonne National Laboratory in \textit{Short-Lived Germanium and Arsenic Fission Activities} \cite{Ste51} as part of the Manhattan Project Technical Series. Samples of uranyl nitrate were irradiated in the thimble of the Heavy-water Pile and the subsequent period of activity was found to be 40~hours. ``Two germanium and two arsenic activities have been found in fission, with the following chain relations: 12h Ge$^{77} \rightarrow$ 40h As$^{77} \rightarrow$ stable Se$^{77}$,...'' The accepted half-life of $^{77}$As is 38.83(5) h. The discovery had been reported in the 1946 Plutonium Project Record \cite{PPR46a} and the result tabulated in reference \cite{PPR46b} but we only recognize the first unclassified publication of 1951 as the equivalent to a refereed paper. Thus, the credit of discovery should go to Arnold and Sugarman also from Argonne National Laboratory who published their observation of $^{77}$As \cite{Arn47} in 1947. However, because Arnold and Sugarman were aware of and had access to the Plutonium Project work we credit the discovery to Steinberg and Engelkemeir. It should also be mentioned that in 1939 Sagane \cite{Sag39} incorrectly identified $^{77}$As as pointed out by Elliott and Deutsch \cite{Ell43}.

\subsection*{$^{78}$As}\vspace{-0.85cm}
Snell discovered $^{78}$As in 1937 at the Radiation Laboratory at the University of California in Berkeley which he announced in \textit{The Radioactive Isotopes of Bromine: Isomeric Forms of Bromine 80} \cite{Sne37}. ``The reaction $^{81}$Br(n,$\alpha$)$^{78}$As resulted from an activation of a large sample of ammonium bromide with (Be+D) neutrons. After a chemical separation, the arsenic fraction showed activity having a decay period of 65$\pm$3~minutes. This activity is new, and it has been attributed to arsenic 78, presumably made by a parallel reaction from the other bromine isotope.'' The accepted half-life value is 90.7(2)~m.

\subsection*{$^{79}$As}\vspace{-0.85cm}
$^{79}$As was first observed by Butement in 1950 at the Atomic Energy Research Establishment in Harwell, England, as reported in \textit{New Radioactive Isotopes Produced by Nuclear Photo-Disintegration} \cite{But50}. $^{79}$As was produced through irradiation of potassium selenate by 23~MeV x-rays from the synchrotron in the photonuclear reaction $^{80}$Se($\gamma$,p) and chemically separated from other resultant isotopes. ``The activity showed a half-life of 9~minutes and a weak residual activity with an apparent half-life of about 31~hours. The latter may be attributed to a mixture of 26.8-hour $^{76}$As and 40-hour $^{77}$As. The yields of the 9-minute and 31-hour activities were approximately equal. The 9-minute arsenic is therefore probably $^{79}$As, decaying by beta-particle emission into $^{79}$Se whose half-life is either very short or very long.'' The measured half-life agrees with the accepted value of 9.01(15)~m.

\subsection*{$^{80}$As}\vspace{-0.85cm}
Ythier and Herrman discovered $^{80}$As in 1954 at the Max-Planck-Institut f\"{u}r Chemie in Mainz, Germany, which was described in \textit{\"{U}ber Schwere Isotope des Arsens} \cite{Yth54}. $^{80}$As was produced by bombarding selenium with fast neutrons. The isotope was produced in the reaction $^{80}$Se(n,p)$^{80}$As. ``Eine neue und sehr intensive Aktivit\"{a}t von T$\approx$36~sec scheint dem Arsen zuzugeh\"{o}ren und k\"{o}nnte eventuell das gesuchte $^{80}$As sein.'' (A new and very intense activity with T$\sim$ 36 s seems to be due to arsenic and could be the searched for $^{80}$As.) The accepted half-life is 15.2(2)~s.

\subsection*{$^{81}$As}\vspace{-0.85cm}
Morinaga \textit{et al}. discovered $^{81}$As in 1960 at Tohoku University in Sendai, Japan, which was reported in \textit{Three New Isotopes, $^{63}$Co, $^{75}$As, $^{81}$As} \cite{Mor60}. Selenium was bombarded with 25~MeV bremsstrahlung in the betatron and produced via the reaction $^{82}$Se($\gamma$,p). ``Besides all previously known activities a very short-lived component with approximately 30~sec was observed...Fig. 6 shows the half-life of this short half-lived activity measured by the plastic scintillator discriminated at 2.3-MeV. From this measurement the half-life was determined to be 32$\pm$2~sec.'' Chemical separations were performed to confirm the activity was due to arsenic. The half-life is consistent with the accepted value of 33.3(8)~s. It should be mentioned that in the same month (February 1960) C. Ythier confirmed the results \cite{Yth60}. C. Ythier had submitted his paper about three months later and was aware of the Morinaga manuscript.

\subsection*{$^{82}$As}\vspace{-0.85cm}
In 1968 at the Institute of Nuclear Sciences in Wellington, New Zealand, Mathew \textit{et al}. discovered $^{82}$As which was reported in \textit{New Isotope $^{82}$As} \cite{Mat68}. The new isotope was produced in the reaction $^{82}$Se(n,p)$^{82}$As where the neutrons were produced by bombarding a tritium target with 0.8 MeV deuterons. ``Two new $\gamma$-activities of energies 655$\pm$0.5~keV and 817$\pm$0.5~keV and half-life 15$\pm$2~s have been produced by irradiation of natural selenium and enriched $^{82}$Se, with 16.4~MeV neutrons. These activities are assigned to the $\beta$$^-$ decay of $^{82}$As formed in the reaction $^{82}$Se(n,p)$^{82}$As.'' A coaxial Ge(Li)detector was used to measure the $\gamma$-ray spectrum. The measured half-life is close to the accepted value of 19.1(5) s.

\subsection*{$^{83-84}$As}\vspace{-0.85cm}
In \textit{Identification of new arsenic isotopes in fission: $^{83}$As and $^{84}$As} del Marmol reported the discovery of $^{83-84}$As at the Centre d'Etude de l'Energie Nucl\'eaire in Belgium in 1968 \cite{Mar68}. Thermal neutrons from a BR2 reactor irradiated a solution of $^{235}$U, $^{76}$As tracer, As$^{+5}$, Sb$^{+5}$ and SeO$_{3}$ dissolved in sulfuric acid. ``A least-squares analysis, weighted for initial bromine activities, gives half-lives of 14.1$\pm$1.1 sec for $^{83}$As and of 5.8$\pm$0.5 sec for $^{84}$As.'' The value for $^{83}$As is included in the average for the accepted half-life of 13.4(3) s and the value for $^{84}$As is close to the accepted half-life of 4.02(3) s.

\subsection*{$^{85}$As}\vspace{-0.85cm}
In 1967 at Mol, Belgium, del Marmol and de Mevergnies were the first to identify $^{85}$As which they reported in \textit{Investigation of delayed neutron precursors of As, Sb and Ge} \cite{Mar67}. $^{85}$As was produced in thermal neutron fission of $^{235}$U and identified via chemical separation where a neutron activity of 2.14~s was observed. ``Regarding mass assignments for this 2.15-sec activity, prospective d.n.p. [delayed neutron precursors] are expected among isotopes within masses 85 and 87; ... However, owing to the absence of any 56-sec neutron activity from the $^{87}$Br grand-daughter, $^{87}$As could be ruled out as being responsible for the 2.15-sec activity,...''. The half-life agrees with the presently accepted value of 2.021(10) s.

\subsection*{$^{86}$As}\vspace{-0.85cm}
 In 1973 Kratz \textit{et al}. were the first to observe and separate $^{86}$As at the Institut f\"{u}r Anorganische und Kernchemie der Universit\"{a}t Mainz in Germany which they reported in \textit{Delayed-Neutrons from Arsenic Isotopes $^{84}$As, $^{85}$As, and $^{86}$As} \cite{Kra73}. $^{86}$As was produced via the thermal-neutron fission of $^{235}$U and subsequent isolation by volatilization of arsenic hydride. ``A new isotope, 0.9$\pm$0.2~sec $^{86}$As, was detected by delayed neutron counting and by following the decay of its most prominent $\gamma$-ray. The mass assignment was verified by milking of 54~sec $^{86}$Br.'' The measured half-life agrees with the accepted value of 0.945(8)~s.

\subsection*{$^{87}$As}\vspace{-0.85cm}
In 1970 Kratz and Herrmann in \textit{Half-Lives, Fission Yields and Neutron Emission Probabilities of $^{87}$Se and $^{88}$Se, and Evidence for $^{87}$As} reported the observation of $^{87}$As at the Institut f\"{u}r Anorganische und Kernchemie der Universit\"{a}t Mainz in Germany, via thermal-neutron fission of $^{235}$U in the Mainz Triga reactor \cite{Kra70}. Neutron activities were measured with $^3$He counting tubes and $\gamma$-ray spectroscopy was measured with a Ge(Li) diode. ``Evidence for the existence of $^{87}$As was found from a slight growth of the $^{87}$Se activity, corresponding to a half-life of 0.6$\pm$0.3~sec and a fractional cumulative yield of 4$\pm$2 per cent for $^{87}$As.'' The measured half-life is consistent with the accepted value of 0.61(12)~s.

\subsection*{$^{88-89}$As}\vspace{-0.85cm}
In 1994 Bernas \textit{et al}. announced the discovery of $^{88-89}$As in \textit{Projectile Fission at Relativistic Velocities: A Novel and Powerful Source of Neutron-Rich Isotopes Well-Suited for In-Flight Isotopic Separation} at GSI in Darmstadt, Germany \cite{Ber94}. A 750~A$\cdot$MeV $^{238}$U beam accelerated by the heavy ion synchrotron SIS impinged on a lead target. ``Reaction products were analyzed with the fragment separator FRS which was operated in the achromatic mode. Energy loss of the separation fragments, which is characteristic for their nuclear charge Z, was measured in a four-stage MUSIC ionization chamber at the exit of the FRS'' which allowed particles to be ``unambiguously identified by their energy-loss and time-of-flight.'' 51 counts of $^{88}$As and 8 counts of $^{89}$As were observed.

\subsection*{$^{90-92}$As}\vspace{-0.85cm}
Bernas \textit{et al}. reported the observation of $^{90-92}$As in \textit{Discovery and Cross-Section Measurement of 58 New Fission Products in Projectile-Fission of 750$\cdot$AMeV $^{238}$U} at GSI in Darmstadt, Germany, in 1997 \cite{Ber97}. The experimental and analysis procedures were the same as in the 1994 experiment with the only difference being that a beryllium target was used instead of a lead target. ``The projectile fission of uranium at relativistic energy impinging on a Be target was investigated with the fragment separator, FRS, in order to produce and identify new isotopes and to measure their production yields.'' 228 counts of $^{90}$As, 37 counts of $^{91}$As and four counts of  $^{92}$As were observed.

\section{Summary}
The discoveries of the isotopes of arsenic have been catalogued and the methods of their production discussed. Only two isotopes had been previously wrongly identified ($^{73}$As and $^{77}$As). The decay of $^{70}$As was undoubtedly observed, but incorrectly attributed to $^{71}$As, two years before the discovery publication we have accepted. The discovery of $^{66}$As was not acknowledged by the authors because of the presence of an unpublished report.
$^{77}$As demonstrates the difficulty in assigning the discovery of isotopes during the Manhattan Project. As a general guideline we consider the unclassified publication of the Plutonium Project Records in 1951 as the relevant publication \cite{PPR51}. However, in the case of $^{77}$As Arnold and Sugarman published their result in 1947 \cite{Arn47} being aware of and having access to the data of their colleagues within the Plutonium Project. Thus, we attribute the discovery of $^{77}$As to Steinberg and Engelkemeir \cite{Ste51}.

\ack

This work was supported by the National Science Foundation under grants No. PHY06-06007 (NSCL) and PHY07-54541 (REU). MH was supported by NSF grant PHY05-55445.


\newpage

\section*{EXPLANATION OF TABLE}\label{sec.eot}
\addcontentsline{toc}{section}{EXPLANATION OF TABLE}

\renewcommand{\arraystretch}{1.0}

\begin{tabular*}{0.95\textwidth}{@{}@{\extracolsep{\fill}}lp{5.5in}@{}}
\textbf{TABLE I.}
	& \textbf{Discovery of Arsenic Isotopes }\\
\\

Isotope & Arsenic isotope \\
First Author & First author of refereed publication \\
Journal & Journal of publication \\
Ref. & Reference \\
Method & Production method used in the discovery: \\
  & FE: fusion evaporation \\
  & LP: light-particle reactions (including neutrons) \\
  & MS: mass spectroscopy \\
  & NF: neutron-induced fission \\
  & PN: photo-nuclear reactions \\
  & PF: projectile fragmentation or projectile fission \\
Laboratory & Laboratory where the experiment was performed\\
Country & Country of laboratory\\
Year & Year of discovery \\
\end{tabular*}
\label{tableI}

\newpage
\datatables

\setlength{\LTleft}{0pt}
\setlength{\LTright}{0pt}


\setlength{\tabcolsep}{0.5\tabcolsep}

\renewcommand{\arraystretch}{1.0}


\begin{longtable}[c]{%
@{}@{\extracolsep{\fill}}r@{\hspace{5\tabcolsep}} llllllll@{}}
\caption[Discovery of Arsenic Isotopes]%
{Discovery of Arsenic isotopes}\\[0pt]
\caption*{\small{See page \pageref{tableI} for Explanation of Tables}}\\
\hline
\\[100pt]
\multicolumn{8}{c}{\textit{This space intentionally left blank}}\\
\endfirsthead
Isotope & First Author & Journal & Ref. & Method & Laboratory & Country & Year \\

$^{64}$As & B. Blank & Phys. Rev. Lett. & Bla95 & PF & GANIL & France &1995 \\
$^{65}$As & M.F. Mohar & Phys. Rev. Lett. & Moh91 & PF & Michigan State & USA &1991 \\
$^{66}$As & D.E. Alburger & Phys. Rev. C & Alb78 & PF & Brookhaven & USA &1978 \\
$^{67}$As & M.J. Murphy & Phys. Rev. C & Mur80 & PF & Argonne & USA &1980 \\
$^{68}$As & T. Paradellis & Nucl. Phys. A & Par71 & LP & McGill & Canada &1971 \\
$^{69}$As & F.D.S. Butement & Phil. Mag. & But55 & LP & Harwell & UK &1955 \\
$^{70}$As & H.H. Hopkins Jr. & Phys. Rev. & Hop50 & LP & Berkeley & USA &1950 \\
$^{71}$As & R. Sagane & Phys. Rev. & Sag39 & LP & Berkeley & USA &1939 \\
$^{72}$As & A.C.G. Mitchell & Phys. Rev. & Mit47 & LP & Indiana & USA &1939 \\
$^{73}$As & D.A. McCown & Phys. Rev. & McC48 & LP & Ohio State & USA &1948 \\
$^{74}$As & R. Sagane & Phys. Rev. & Sag38 & LP & Berkeley & USA &1938 \\
$^{75}$As & F.W. Aston & Phil. Mag. & Ast20 & MS & Cavendish & UK &1920 \\
$^{76}$As & E. Amaldi & Ric. Scientifica & Ama34 & NF & Rome & Italy &1934 \\
$^{77}$As & E.P. Steinberg & Nat. Nucl. Ener. Ser. & Ste51 & NF & Argonne & USA &1951 \\
$^{78}$As & A.H. Snell & Phys. Rev. & Sne37 & LP & Berkeley & USA &1937 \\
$^{79}$As & F.D.S. Butement & Proc. Roy. Soc. & But50 & PN & Harwell & UK &1950 \\
$^{80}$As & C. Ythier & Z. Elektrochem & Yth54 & LP & Mainz & Germany &1954 \\
$^{81}$As & H. Morinaga & J. Phys. Soc. Japan & Mor60 & PN & Tohoku & Japan &1960 \\
$^{82}$As & P.J. Mathew & Phys. Lett. B & Mat68 & LP & Wellington & New Zealand &1968 \\
$^{83}$As & P. del Marmol & J. Inorg. Nucl. Chem. & Mar68 & NF & Mol & Belgium &1968 \\
$^{84}$As & P. del Marmol & J. Inorg. Nucl. Chem. & Mar68 & NF & Mol & Belgium &1968 \\
$^{85}$As & P. del Marmol & J. Inorg. Nucl. Chem. & Mar67 & NF & Mol & Belgium &1967 \\
$^{86}$As & J.V. Kratz & J. Inorg. Nucl. Chem. & Kra73 & NF & Mainz & Germany &1973 \\
$^{87}$As & J.V. Kratz & J. Inorg. Nucl. Chem. & Kra70 & NF & Mainz & Germany &1970 \\
$^{88}$As & M. Bernas & Phys. Lett. B & Ber94 & PF & Darmstadt & Germany &1994 \\
$^{89}$As & M. Bernas & Phys. Lett. B & Ber94 & PF & Darmstadt & Germany &1994 \\
$^{90}$As & M. Bernas & Phys. Lett. B & Ber97 & PF & Darmstadt & Germany &1997 \\
$^{91}$As & M. Bernas & Phys. Lett. B & Ber97 & PF & Darmstadt & Germany &1997 \\
$^{92}$As & M. Bernas & Phys. Lett. B & Ber97 & PF & Darmstadt & Germany &1997 \\

\end{longtable}

\newpage


\normalsize

\begin{theDTbibliography}{1956He83}

\bibitem[Alb78]{Alb78t} D.E. Alburger, Phys. Rev. C {\bf 18}, 1875 (1978)
\bibitem[Ama34]{Ama34t} E. Amaldi, O. D'Agostino, E. Fermi, F. Rasetti, and E. Segr\`{e}, Ric. Scientifica {\bf 5}, 21 (1934)
\bibitem[Ast20]{Ast20t} F.W. Aston, Phil. Mag. {\bf 40}, 632 (1920)
\bibitem[Ber94]{Ber94t} M. Bernas \textit{et al}., Phys. Lett. B {\bf 321}, 19 (1994)
\bibitem[Ber97]{Ber97t} M. Bernas \textit{et al}., Phys. Lett. B {\bf 415}, 111 (1997)
\bibitem[Bla95]{Bla95t} B. Blank, \textit{et al}., Phys. Rev. Lett. {\bf 74}, 4611 (1995)
\bibitem[But50]{But50t} F.D.S. Butement, Proc. Roy. Soc.(London) {\bf 64A}, 395 (1951)
\bibitem[But55]{But55t} F.D.S. Butement and E.G. Prout, Phil. Mag. {\bf 46}, 357 (1955)
\bibitem[Hop50]{Hop50t} H.H. Hopkins Jr., Phys. Rev. {\bf 77}, 717 (1950)
\bibitem[Kra70]{Kra70t} J.V. Kratz and G. Herrmann, J. Inorg. Nucl. Chem {\bf 32}, 3713 (1970)
\bibitem[Kra73]{Kra73t} J.V. Kratz, H. Franz, and G. Herrmann, J. Inorg. Nucl. Chem {\bf 35}, 1407 (1973)
\bibitem[Mar67]{Mar67t} P. del Marmol and M. N\`eve de M\'evergnies, J. Inorg. Nucl. Chem {\bf 29}, 273 (1967)
\bibitem[Mar68]{Mar68t} P. del Marmol, J. Inorg. Nucl. Chem {\bf 30}, 2873 (1968)
\bibitem[Mat68]{Mat68t} P.J. Mathew and G.J. McCallum, Phys. Lett. {\bf 28B}, 106 (1968)
\bibitem[McC48]{McC48t} D.A. McCown, L.L. Woodward, and M.L. Pool, Phys. Rev. {\bf 74}, 1315 (1948)
\bibitem[Mit47]{Mit47t} A.C.G. Mitchell, E.T. Jurney, and M. Ramsey, Phys. Rev. {\bf 71}, 825 (1947)
\bibitem[Moh91]{Moh91t} M.F. Mohar, D. Bazin, W. Benenson, D.J. Morrissey, N.A. Orr, B.M. Sherrill, D. Swan, J.A. Winger, A.C. Mueller, and D. Guillemaud-Mueller, Phys. Rev. Lett. {\bf 66}, 1571 (1991)
\bibitem[Mor60]{Mor60t} H. Morinaga, T. Kuroyanagi, H. Mitsui, and K. Shoda, J. Phys. Soc. Japan {\bf 15}, 213 (1960)
\bibitem[Mur80]{Mur80t} M.J. Murphy, C.N. Davids, and E.B. Norman, Phys. Rev. C {\bf 22}, 2204 (1980)
\bibitem[Par71]{Par71t} T. Paradellis, A. Houdayer, and S.K. Mark, Nucl. Phys. A {\bf 174}, 617 (1971)
\bibitem[Sag38]{Sag38t} R. Sagane, S. Kojima, and M. Ikawa, Phys. Rev. {\bf 54}, 149 (1938)
\bibitem[Sag39]{Sag39t} R. Sagane, Phys. Rev. {\bf 55}, 31 (1939)
\bibitem[Sne37]{Sne37t} A.H. Snell, Phys. Rev. {\bf 52}, 1007 (1937)
\bibitem[Ste51]{Ste51t} E.P Steinberg and D.W. Engelkemeir, {\it Radiochemical Studies: The Fission Products}, Paper 54, p. 566, National Nuclear Energy Series IV, 9, (McGraw-Hill, New York 1951)
\bibitem[Yth54]{Yth54t} C. Ythier and G. Herrmann, Z. Elektrochem. {\bf 58}, 1630 (1954)

\end{theDTbibliography}

\end{document}